\documentclass{iaus}
\usepackage{graphicx}
\usepackage{subfigure}

\newcommand{\Hb}{\ifmmode {\rm H}\beta \else H$\beta$\fi}

\title[Beware of fake AGNs] 
{Beware of fake AGNs}

\author[G. Stasi\'nska]   
{G. Stasi\'nska$^1$,
        N. Vale Asari$^{1,2}$,
            R. Cid Fernandes$^2$\\
                 \small{for the SEAGal collaboration (Semi-Empirical Analysis of GALaxies)}}
\affiliation{$^1$LUTH, Observatoire de Paris, CNRS, Universit\'e Paris Diderot; Meudon, France\\
$^2$Dpto de F\'{\i}sica - CFM - Universidade Federal de Santa Catarina,
	 Florian\'opolis, SC, Brazil
}

\pubyear{2009}
\volume{267}  
\pagerange{ }
\jname{Co-evolution of central
black holes and galaxies: feeding and feedback}
\editors{Brad Peterson, Rachel Somerville, \& Thaisa
Storchi-Bergmann eds.}
\begin{document}

\maketitle

\begin{abstract}
In the BPT diagram, the distribution of the emission-line galaxies from the Sloan Digital Sky Survey (SDSS) evokes the wings of a seagull. Traditionally, galaxies in the right wing are considered to host AGNs.  
Our study of the stellar populations of SDSS galaxies showed that $\sim$ 1/4 of galaxies thought to host LINERS are in fact ``retired galaxies'', i.e. galaxies that stopped forming stars and are ionized by hot post-AGB stars and white dwarfs (Stasi\'nska et al. 2008). 
When including the  galaxies that lack some of the lines needed to place them in the BPT diagram the fraction of retired galaxies is even larger (Cid Fernandes et al., 2009, arXiv:0912.1376).
\end{abstract}

\begin{figure*} [h]
  \begin{center}
\subfigure{\includegraphics[width=7cm]{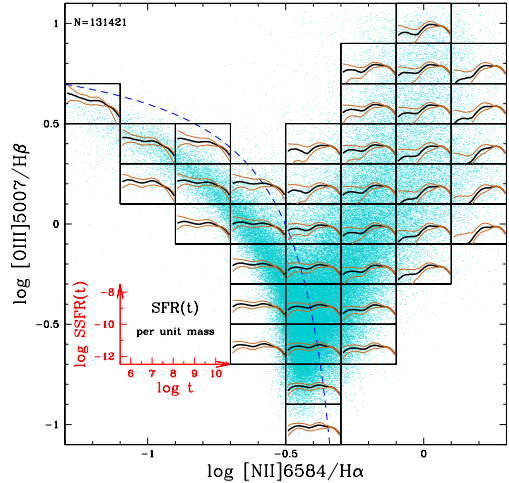}} 
    \end{center}
  \caption{The BPT diagram for 131421 galaxies in the SDSS. The dashed line separates pure star forming galaxies from the rest (Stasi\'nska et al. 2006). Superimposed are Specific Star Formation Histories obtained with the stellar population synthesis code  {\sc starlight} (Cid Fernandes et al. 2005). Clearly, Seyfert galaxies (upper branch of the right wing) still form stars while LINERs (lower branch) do not.  Photoionization models using the radiation from the stellar populations in the LINER region are able to cover the entire BPT plane (Stasi\'nska et al 2008). The  seagull shape is the result of selection effects.
}
    \label{fig:image}
\end{figure*}





\begin{thebibliography}{}
\bibitem[\protect\citeauthoryear{Cid Fernandes et 
al.}{2005}]{2005MNRAS.358..363C} Cid Fernandes R., Mateus A., Sodr{\'e} L., et al., 2005, MNRAS, 358, 363 


\bibitem[\protect\citeauthoryear{Stasi{\'n}ska et 
al.}{2006}]{2006MNRAS.371..972S} Stasi{\'n}ska G., Cid Fernandes R., Mateus 
A., et al., 2006, MNRAS, 371, 972 

\bibitem[\protect\citeauthoryear{Stasi{\'n}ska et 
al.}{2008}]{2008MNRAS.391L..29S} Stasi{\'n}ska G., Vale Asari N., Cid 
Fernandes R., et al., 2008, MNRAS, 391, L29 
\end{thebibliography}
\end{document}